\documentstyle[aps,preprint]{revtex}
\begin{document}
\draft
\title{ Duality picture between antiferromagnetism and 
d-wave superconductivity in $t-J$ model at two dimensions
}
\author{Tai Kai Ng}
\address{Dept. of Physics, HKUST, Kowloon, Hong Kong}
\date{ \today }
\maketitle
\begin{abstract}
  We show in this paper an interesting relation between elementary
and topological excitations in the antiferromagnetic and d-wave 
superconducting phases of the $t-J$ model at two dimensions. The
topological spin and charge excitations in one phase have the same 
dynamics as elementary excitations in the other phase, except the 
appearance of energy gaps. Moreover, the transition from one phase to 
another can be described as a quantum disordering transition associated 
with the topological excitations. Based on the above picture, a 
plausible phase diagram of $t-J$ model is constructed.
\end{abstract} 

\pacs{74.20.Mn,74.72.-h,75.10.Jm}

\narrowtext

\section{Introduction}
  It is now commonly believed that a large part of the phase diagram 
of high-$T_c$ superconductors and corresponding complicated
phenomenologies are results of subtle interplay between antiferromagnetism
and (d-wave) superconductivity in these materials. There exists now many 
theoretical attempts to understand the relationship between the
antiferromagnetic and superconducting phases in the cuprates. For example, 
Pines {\em et.al.} have tried to interpret the d-wave superconducting phase 
as an almost antiferromagnetic Fermi liquid\cite{pines}, with 
superconductivity being driven by antiferromagnetic spinwave fluctuations. 
A somewhat related approach was also employed by Sarker\cite{sarker} using 
the slave-fermion approach to the $t-J$ model\cite{sfmft}, where he
proposed that d-wave superconductivity arises in the hole-driven quantum
disordered phase of antiferromagnets as a result of formation of spin
resonant-valence-bonds (RVB) and spin-charge recombination. Another 
completely different approach was proposed by Zhang 
{\em et.al.}\cite{zhang} where they proposed that antiferromagnetism and 
d-wave superconductivity are related by an hidden SO(5) symmetry in the
system\cite{zhang}. All these proposals have met criticisms of one form 
or other and there is no universally accepted theory at present.

   In this paper we shall propose a duality relation between
antiferromagnetism and d-wave superconductivity in the $t-J$ model. 
In existing treatments of $t-J$ model, antiferromagnetism and (d-wave)
superconductivity appear separately in very different ways of treating the
problem. The antiferromagnetic (or spiral) state is best described by the 
slave-fermion mean-field theory (SFMFT)\cite{sfmft} where the spins 
are represented by bosons and holes by fermions, whereas d-wave
superconductivity is described by the slave-boson mean-field theory 
(SBMFT)\cite{sbmft} where spins are represented by fermions and holes 
by bosons. The properties of the ground states and the corresponding
elementary excitations are very different in the two mean-field theories. 
In SFMFT, the elementary excitations are bosonic spins (spinwaves) and 
fermionic holes. d-wave superconductivity and Fermi surface satisfying 
the Luttinger Theorem are absent in the mean-field treatment. On the 
other hand, SBMFT describes the d-wave superconducting state, spin-gap 
phase, and Fermi-surfaces fairly satisfactory in the underdoped to
optimally-doped regime of high-$T_c$ cuprates, but fails to describe 
correctly the antiferromagnetic (or spiral) state. The elementary 
excitations in SBMFT are fermionic spins and bosonic holes. In this 
paper, we shall propose that dispite the huge differences in the structure 
of the mean-field theories, an intimate relation exists between the two 
phases of the $t-J$ model. The elementary excitations in one mean-field 
theory can be viewed as topological excitations in the other mean-field 
theory and the mean-field ground state of one theory can be viewed as 
a quantum disordered state formed by "condensation" of topological 
excitations obtained in the other mean-field theory. Based on this 
physical picture and with some self-consistency requirements, we shall 
construct a plausible $\delta-T$ phase diagram for the $t-J$ model. The
organization of our paper is as follows. In section II we shall discuss 
the construction and dynamics of the topological spin and charge 
excitations in SFMFT where we shall show that the topological excitations 
in SFMFT have same statistics and dynamics as elementary excitations in 
the SBMFT except that they are gapped in the antiferromagnetic phase. In 
section III We shall discuss the corresponding case of topological 
excitations in SBMFT where we shall show that similar relations exist 
between the topological excitations in SBMFT and elementary excitations 
in SFMFT. With these results we shall argue in section IV that the ground 
state described by one mean-field theory can be viewed as a quantum 
disordered state of topological excitations obtained in the other 
mean-field theory. Based on this physical picture and with some
self-consistency requirements, we shall propose a plausible $\delta-T$ 
phase diagram of the $t-J$ model. Our results are summarized in section 
V where some experimental consequences of our theory will be discussed 
and a physical picture of the duality relation based on 
Resonant-Valence-Bond (RVB) picture will be presented.

  Before we proceed further we first clarify a notation. In the 
following we shall call the topological spin and charge excitations 
'spinons' and 'holons', respectively to distinguish them from 
corresponding elementary excitations in mean-field theories. Notice 
that there are two different kinds of spinons and holons, corresponding 
to two different mean-field theories.  

\section{topological excitations in slave-fermion mean-field theory}
    In SFMFT, the spins are represented by Schwinger boson operators
$\bar{Z}_{i\sigma}, Z_{i\sigma}$ whereas holes are represented by slave 
fermion operators $f^+_i, f_i$. In terms of these operators the mean-field
Hamiltonian has the following form, $H_{MF}=H_{mf}^s+H_{mf}^h$, 
where\cite{sfmft}
\begin{mathletters}
\label{hmf}
\begin{eqnarray}
\label{hs}
H_{mf}^s & = & -{J\over2}\sum_{<i,j>}\left(\Delta^*(Z_{i\uparrow}Z_{j\downarrow}
-Z_{i\downarrow}Z_{j\uparrow}) + H.C.\right)  \\   \nonumber
& & +\sum_{<i,j>,\sigma}\left(({J\over2}\chi^*_{\nu}+tF^*_{\nu})
\bar{Z}_{i\sigma}Z_{j\sigma}+H.C.\right)+\sum_{i}\lambda
\left(\sum_{\sigma}(\bar{Z}_{i\sigma}Z_{i\sigma}-(2S-\delta)\right),
\end{eqnarray}
and
\begin{equation}
\label{hh}
H_{mf}^h=\sum_{<i,j>}\left(t\chi_{\nu}f^+_jf_i+ H.C.\right)
-\sum_{i}\mu{f}^+_if_i,
\end{equation}
\end{mathletters}
where $<i,j>$ are nearest neighbor pairs of sites on a square lattice, 
$j=i+\nu$ where $\nu=\pm\hat{x},\pm\hat{y}$. We shall put site $i$'s on 
A-sublattice and site $j$'s on B-sublattice in all our following
discussions. $S=1/2$ is the spin magnitude in $t-J$ model. $\Delta$,
$\chi_{\nu}$, $F_{\nu}$ and $\lambda$ are mean-field parameters 
determined by the mean-field equations $\Delta=<Z_{i\uparrow}Z_{j\downarrow}-Z_{i\downarrow}Z_{j\uparrow}>$, 
$\chi_{\nu}=\sum_{\sigma}<\bar{Z}_{i\sigma}Z_{j\sigma}^B>$,
$F_{\nu}=<f^+_jf_i>$, $<\bar{Z}_{i\uparrow}Z_{i\uparrow}>+
<\bar{Z}_{i\downarrow}Z_{i\downarrow}>=2S-\delta$ and $<f^+_if_i>=
\delta$. We shall choose a gauge where the mean-field parameters 
$\Delta$ have s-symmetry, and $\chi_{\nu}$ and $F_{\nu}$ have 
p-symmetry $\chi(F)_{-\nu}=-\chi(F)_{\nu}$ in the spiral 
states\cite{sfmft}. Notice that at two dimensions, long-ranged (spiral)
antiferromagnetic order exists at zero temperature in SFMFT at small 
doping, corresponding to Bose-condensation $<Z>\neq0$ in mean-field 
theory. We shall consider finite temperature $T\neq0$ and $<Z>=0$ in 
the following. The effect of Bose condensation will be addressed at the 
end of the section.

   To look for topological excitations in SFMFT, we notice that
the structure of the mean-field theory resembles very much the BCS
theory for superconductivity, except that spin-pairs of bosons replace
the electron (fermion) Cooper pairs in BCS theory. The resemblance of 
the two theories leads us to study {\em vortex} excitations in SFMFT, 
since vortices are stable topological excitations in BCS theory at
two dimensions. In BCS theory, a vortex located at $\vec{r}=0$
is a solution of the BCS mean-field equation, where the order
parameter $\Delta_{BCS}(\vec{r})$ has a form 
\[
\Delta_{BCS}(\vec{r})=f(r)e^{i\theta}, \]
in a polar coordinate, where $f(r)$ is real and positive. To
minimize energy, a magnetix flux of $\pi$-flux quanta is trapped
in the vortex core. The vortex solution in SFMFT has the same 
structure, except that the BCS order parameter $\Delta_{BCS}$ is 
replaced by the Schwinger boson order parameter $\Delta_{ij}$ and 
the vector potential $\vec{A}$ does not represent the physical 
magnetic field, but is a fictitious gauge field arising from phase 
fluctuations of order parameter $\Delta_{ij}$'s\cite{rs,ng1}. The
existence and stability of the vortex solution in Heisenberg model
was demonstrated in Ref.\cite{ng1,ng2} in an effective
Ginsburg-Landau theory. These vortex solutions are bosonic,
$S=0$ topological excitations in SFMFT\cite{ng1} (see fig.1a).

  To construct topological {\em spin} excitations (spinons) we note, 
that like vortices in superconductors where electronic bound states 
often exist inside the vortex core, bosonic (spin) bound states may 
exist inside vortices in SFMFT. In particular, we have argued that 
in the zero doping limit, for a vortex centered at a lattice site 
(fig.1b), a bound state with one boson must be formed at the vortex 
center because of constraint that there is always one spin per site 
in the Heisenberg model\cite{ng1,ng2}. In particular, because of 
statistics transmutation associated with binding quantum particle 
to flux-tubes of $\pi$-flux quantum in two dimensions\cite{wz}, the 
resulting excitation is a spin-$1/2$ fermion\cite{ng1,ng2,rc}. 

   In the presence of holes, a new type of vortex solution with
vortex center located at a lattice site may form. In this case,
one may remove a spin from the center of vortex, leaving a hole (empty
site) there (fig.1c). The resulting object has charge one, spin 
zero and is a topological hole excitation (holon). Because of 
statistics transmutation effect associated with binding quantum 
particles to $\pi$-flux quantum, the resulting holon is a 
boson\cite{ng1}. 

   It is helpful to view the vortex excitations from a
"wavefunction" picture\cite{rc}. First we consider the zero-doping
limit of Heisenberg model. In this limit, the slave-fermion mean-field
theory with $<Z>=0$ can be understood as the mean-field description 
of a short-ranged RVB (resonant valence bond) state\cite{rc,liang}
where the wavefunction is made up of short-ranged spin-singlet pairs 
with the two spins of the singlet resting on opposite sublattices. 
The relative phases between the spin singlets are fixed by the 
Marshall sign rule, and the total wavefunction represents a 
Bose-condensate of spin singlet pairs, just like a BCS superconducting 
state can be viewed as a Bose-condensed state of Cooper pairs. The 
coherent phase structure of the ground state wavefunction allows us 
to construct topological excitations (vortices) in this condensate 
where the phases of the spin singlets change by $2\pi$ when going 
around the center of a vortex\cite{rc}. Notice that although SFMFT 
does not enforce the requirement of no double occupancy in the 
$t-J$ model rigorously, the correct phase structure of the 
wavefunction (Marshall sign rule) is kept. Therefore we expect 
that the qualitative properties of vortex excitations are correct in 
SFMFT. In the spiral states the consideration is similar, except that the 
two spins in a spin-singlet pair may occupy the {\em same} sublattice 
in the spiral states\cite{sfmft} and the phases are not given 
by Marshall sign rule but are nevertheless fixed\cite{sfmft}. Similar
consideration can also be applied to construct vortex excitations 
in SBMFT.

  To study the dynamics of holons and spinons, we shall consider
the continuum limit of SFMFT and derive an effective action for
vortices in the limit of small hole concentration $\delta$. An 
important difference between SFMFT and BCS theory is that the mean-field 
Hamiltonian \ (\ref{hmf}) breaks translational symmetry of the
Heisenberg model by one lattice site. Correspondingly, to describe
the dynamics in SFMFT correctly we must keep two lattice sites per 
unit cell in SFMFT\cite{rs} and the fluctuations of the order 
paramter $\Delta_{ij}$'s are described in general by two amplitude and 
two phase (uniform and staggered) fields in the continuum 
limit\cite{rs,ng1}, i.e.
\begin{equation}
\label{delta}
\Delta_{i,i\pm\nu}={1\over2}\left(\phi(i\pm{\nu\over2})+
q_{\pm\nu}(i\pm{\nu\over2})\right)e^{i\left[\int^{i\pm{\nu\over2}}
2\vec{A}.d\vec{x}+A^s_{\pm\nu}(i\pm{\nu\over2})\right]},
\end{equation}
where $q_{-\nu}=-q_{\nu}$, $A^s_{-\nu}=-A^s_{\nu}$ are 'staggered' 
components of the amplitude and phase fluctuations of $\Delta$, 
respectively, $\phi$ and $\int^x\vec{A}.d\vec{x}'$ are the corresponding 
'uniform' components. Correspondingly, the constraint field $\lambda$
is also separated into 'uniform' and 'staggered' components $\lambda^u$ 
and $\lambda^s$. In momentum space ($\vec{k}\neq0$),
\begin{equation}
\label{a0}
\lambda^{u(s)}(\vec{k})={1\over2}
[\lambda^A(\vec{k})+(-)\lambda^B(\vec{k})],
\end{equation}
where $A(B)$ are sublattice indices. The effective Ginsburg-Landau 
action for the continuum field variables is derived in Refs.\cite{rs}
and Ref.\cite{ng1} in the zero hole concentration $\delta=0$ limit. We 
obtain to order $O(m^0)$ ($m$ is the mass gap for spinwave excitations 
in SFMFT which is very small at low temperature ($m\sim{J}e^{-JS/T}$)),
$\lambda^u\sim2\phi$, and $S_{eff}=S_{u}(0)+S_{s}$, where
\begin{equation}
\label{glu}
S_{u}(0)\sim{J\over2}\int{d}\tau\int{d}^2x\left((2a_1-4(2S+1))\phi+2\phi^2
+(2S+1)\phi(2\vec{A})^2\right),
\end{equation}
describes the fluctuations of the 'uniform' fields. $\vec{A}=(A_x,A_y)$,
$S=1/2$ and $a_1<4$ is a numerical constant. Notice that in general a
term $\sim{F}_{\mu\nu}^2$ where $F_{\mu\nu}=\partial_{\mu}A_{\nu}
-\partial_{\nu}A_{\mu}$ also exists in $S_u(0)$ which we have not
included in Eq.\ (\ref{glu}). We find that the 'uniform' gauge field 
$\vec{A}$ acquires a gap $\sim4(2S+1)\phi$ (Meissner effect) 
as a result of nonzero $<\Delta_{ij}>$ in SFMFT. The existence of stable 
vortex solution in SFMFT is tied with the existence of Meissner effect 
as in usual BCS superconductors\cite{ng1,ng2}. Notice that we have 
choosen the London gauge in deriving the action $S_{u}$ which breaks 
the gauge symmetry $\Delta_{i,i\pm\nu}\rightarrow\Delta_{i,i\pm\nu}
e^{i\theta}$, $Z_{i\sigma}\rightarrow{Z}_{i\sigma}e^{i\theta/2}$, and
$Z_{i+\nu\sigma}\rightarrow{Z}_{i+\nu\sigma}e^{i\theta/2}$. The 
correct gauge transformation property can be recovered by replacing
$2\vec{A}\rightarrow2\vec{A}-\nabla\theta$ in $S_{u}$. Correspondingly,
we also have
\begin{equation}
\label{gls}
S_{s}\sim{J\over2}\int{d}\tau\int{d}^2x\left((1-{2b_1\over\phi})
(q_{\mu})^2+{1\over{e}^2}F_{\mu\nu}^{(s)2}+2ic_1F_{\mu\tau}^{(s)}
q_{\mu}\right),
\end{equation}
where $e^2\sim{m}$, $b_1$ and $c_1$ are constants of order O(1).
$F_{\mu\nu}^{(s)}=\partial_{\mu}A^s_{\nu}-\partial_{\nu}A^s_{\mu}$ where
$\mu=\hat{x},\hat{y},\tau$ is a space-time index. The time-component
of staggered gauge field is given by $A^s_{\tau}=\lambda^s$. It turns 
out that the staggered gauge field is not crucial to formation of
vortices but affects the vortex dynamics\cite{ng2}. Since we shall 
treat the dynamics of vortices only at a phenomenological level in this 
paper, we shall concentrate at $S_u$ and ignore $S_s$ in the following.  

  In the presence of holes additional terms appear in the action. In the
continuum limit and for small hole concentration $\delta$, we have
$S_u={S}_u(\delta)+S_h$, where $S_u(\delta)$ is obtained 
from $S_u(0)$ by replacing $2S\rightarrow2S-\delta$ in 
Eq.\ (\ref{glu}), and
\begin{equation}
\label{glh}
S_h=-{t^2\over{J}}\int{d}\tau\int{d}^2x\left
(d_{\alpha}{n_h^2\over{\phi}}\right),
\end{equation}
where $n_h(\vec{x},t)$ is the concentration of holes at space-time 
position $\vec{x},t$, and $d_{\alpha}$ is a constant of order one
which depends on the type of spiral state being formed\cite{sfmft}. 
Gradient terms in $n_h$ are being neglected.

  To derive the dynamics for vortices (spinons and holons), we first
minimize $S_u$ with respect to the $\phi$ field, obtaining for the
ground state
\[
\phi_0\sim{1\over2}\left(2(2S+1-\delta)-a_1-{4d_{\alpha}t^2\delta^2
\over(2(2S+1)-a_1)^2J^2}\right) +O(\delta^3),
\]
in the $\delta{t}<<J$ limit. The dynamics of the 'uniform' phase  
fluctuations can be obtained by expanding $S_u$ around $\phi_0$. We obtain
\begin{mathletters}
\label{spsf}
\begin{equation}
\label{glgauge}
S_{phase}={J\over2}\int{d}\tau\int{d}^2x\left(\xi_x(\nabla\theta-
2\vec{A})^2+\xi_{\tau}(\partial_{\tau}\theta-2A_{\tau})^2\right),
\end{equation} 
where $A_{\tau}=\lambda_u\sim2(\phi-\phi_0)$, and
\begin{equation}
\label{xis}
\xi_x=(2-\delta)\phi_0,\;\;\;\; \xi_{\tau}={1\over2}(1-{d_{\alpha}
\over\phi_0^3}{t^2\delta^2\over{J}^2}).
\end{equation}
\end{mathletters}
We have set $S=1/2$ in writing down $\xi_x$ and have inserted the
$\partial_{\mu}\theta$ factors back to recover gauge invariance in
$S_{phase}$. 

  Next we introduce vortex 3-current $\vec{j}^v$ in the boson phase field,
$\partial_{\mu}\theta\rightarrow\partial_{\mu}\theta
+\partial_{\mu}\theta'$, where $\theta'$ is multivalued and $j^v_{\mu}
=\epsilon_{\mu\nu\lambda}\partial_{\nu}\partial_{\lambda}\theta'\neq0$.
In the London limit where we treat $\phi_0$ as constant a duality
transformation can be performed where we can integrate out the
$\theta$ field to obtain an effective action $S_{v}$ for
vortices\cite{fl},
\begin{mathletters}
\label{sv}
\begin{equation}
\label{sv1}
S_{v}={J\over2}\int{d}\tau\int{d}^2x\left({1\over\xi_x}|(\nabla\times
\vec{a})_x|^2+{1\over\xi_{\tau}}|((\nabla\times\vec{a})_{\tau}|^2
+2i\vec{a}.(\vec{j}^v-2\nabla\times\vec{A})\right)+S^{(0)}_v
\end{equation}
where $\nabla\times\vec{a}\sim(\nabla\theta'-\vec{A})$, $(\nabla
\times\vec{a})_x$ and $(\nabla\times\vec{a})_{\tau}$ are the spacial
and temporal part of $\nabla\times\vec{a}$, respectively. $S^{(0)}_v$
denotes an additional contribution to the vortex action from  
vortex core, $S^{(0)}_v\sim\{$(energy needed to create vortex core, 
$\epsilon_v$)$\times$(length of vortex trajectory in space-time)$\}$. 
For $N$ vortices,
\begin{equation}
\label{sv2}
S^{(0)}_v\sim{\epsilon}_v\sum_{i=1}^N\int{d}l_i=\epsilon_v
\sum_{i=1}^N\int{d}\tau\sqrt{1+{1\over{c_v}^2}
({d\vec{x_i}\over{d}\tau})^2},
\end{equation}
\end{mathletters}
where $\vec{x}_i(\tau)$ represents the trajectory of the $i$th vortex
in Euclidean space-time, $\epsilon_v\sim\xi_xJ$ and $c_v\sim\sqrt{\xi_x/\xi_{\tau}}$. We shall first consider $S^{(0)}_v$ 
in the following.

  Minimizing $S^{(0)}_v$ at real time it is easy to see that 
$S^{(0)}_v$ describes relativistic particles with energy 
$E=\gamma{\epsilon_v}$, where $\gamma=1/\sqrt{1-{\vec{v}^2
\over{c}_v^2}}$ and $\vec{v}$ is the vortex velocity. In the absence 
of $\nabla\times\vec{A}$ term the particles carry 'charge' (vorticity) 
and interact with each other through an effective U(1) gauge field 
$\vec{a}$. For bosonic vortices the corresponding quantum field theory 
is a relativistic theory of scalar electrodynamics with charged
bosons(vortices)\cite{fl}. In the presence of trapped magnetic flux 
inside vortex core $(\nabla\times(2\vec{A})-\vec{j}_v\sim0)$, the electric
3-current $\vec{j}_v$ is screened and the bosons decoupled from the
gauge field $\vec{a}$. The resulting theory is a relativistic theory of 
bosons with short-ranged interactions, as is in the case of vortices in 
usual superconductors.

  To derive dynamics for the spinons we assume that 
once the bosonic spins are bound to the vortex core, their spacial
degree of freedom is quenched and the only modifications to the
pure vortex action \ (\ref{sv}) are: (1)the vortices now
carry spin indices $m=-1/2,1/2$, and there are two spin
component of vortices, and (2)vortices become fermions because
of statistics transmutation. In particular, since the vortex 
action \ (\ref{sv}) is Lorentz invariant, the
dynamics of spinons must be described by $(2+1)d$-relativistic 
field theories of fermions. To proceed further we examine the 
symmetry constraints imposed by the SFMFT. Since there are 
two lattice sites per unit cell in SFMFT, there are also two 
quantum fields $\psi^{A}_{\sigma}$ and $\psi^{B}_{\sigma}$ in 
the continuum theory, representing spin-$\sigma$ vortices 
centered at $A$- and $B$- sublattice sites, respectively\cite{ng1}. 
Under reflection or rotation by $\pi/2$ around center of a square 
plaquette, the $A$- and $B$- sublattices are interchanged and 
correspondingly also the $\psi^{A}_{\sigma}$ and $\psi^{B}_{\sigma}$
fields. Notice that we have considered finite temperature where 
$<Z>=0$ in our discussion and correspondingly parity 
(space-time reflection) symmetry is unbroken. The spinons
are also coupled to gauge fields $A_{\mu}$ and $A^s_{\mu}$
through the bound spins at the center of vortices. To describe these  
couplings, the quantum fields $\psi^{A}_{\sigma}$ and 
$\psi^{B}_{\sigma}$ must also carry 'charge' and are complex\cite{ng2}. 

 The simplest relativistic quantum field theory for fermions which
satisfies these kinematic constraints and respects parity is a theory
of Dirac fermions in $(2+1)d$\cite{ng2,qft}. In the presence of spin degrees
of freedom $\sigma$ there are more than one possible representation of 
Dirac fermions. For later convenience we shall consider a representation 
which corresponds to the quasi-particle excitations around the four nodes 
of a d-wave superconductor in the following. Notice that without a 
complete microscopic theory for spinons, we cannot determine with 
certainty their correct representation. 

 We introduce two four-component spinor field $\Psi_{i}, i=1,2$, where
\begin{equation}
\label{dspinor}
\Psi_{i}(x)=\left(\begin{array}{r}
\psi^{A}_{i\uparrow}(x)  \\
\psi^{B+}_{i\downarrow}(x)  \\
\psi^B_{i\uparrow}(x)    \\
\psi^{A+}_{i\downarrow}(x)  
\end{array}\right).
\end{equation}
  In terms of $\Psi_{i}$ the effective Lagrangian is 
\begin{mathletters}
\label{lspinon}
\begin{equation}
\label{lspinon1}
L_{eff}^s=\sum_{i=1,2}\Psi_i^+(x){\partial\over
\partial\tau}\Psi_i(x)-H_{spinon}, 
\end{equation}
where
\begin{equation}
\label{hspinon}
H_{spinon}=\Psi^+_1\left[\tau^zic_s^{(sf)}\partial_x+(\tau^++\tau^-)
ic_s^{(sf)}\partial_y+m_s^{(sf)}\hat{I}\right]\Psi_1
+(1\leftrightarrow2; x\leftrightarrow{y}),
\end{equation}
\end{mathletters}
where $m_s^{(sf)}\sim\epsilon_v$ and $c_s^{(sf)}\sim{c}_v$.
For d-wave superconductors, $m_s^{(sf)}=0$ and $i=1,2$ represents 
quasi-particle excitations around the Dirac pockets located at 
$\vec{k}=\pm(k_o,k_o)$ and $\vec{k}=\pm(k_o,-k_o)$, 
respectively\cite{sbmft,nodon}. 

   The dynamics of holons can be considered similarly. Assuming that
the spacial degrees of freedom are quenched once a hole is bound 
to the vortex core, the holons are described by relativistic quantum
field theory of charged bosons. As in the case of spinons there are two 
quantum fields $\pi^{(A)}$ and $\pi^{(B)}$ in the continuum theory, 
representing holons centered at $A-$ and $B-$ sublattices. To respect 
parity, we construct quantum fields $\pi^{(+(-))}=
\pi^{(A)}\pm\pi^{(B)}$ which are eigenstates of $\pi/2$ rotation and 
reflection with eigenvalues $\pm1$. We shall assume in the following
that the low energy dynamics of the holons are described by the
$\pi^{(+)}$ field, and with usual relativistic scalar dynamics
\begin{equation}
\label{leffh}
L_{eff}^{h}=|\partial_{\tau}\pi^{(+)}|^2+|c_h^{(sf)}\nabla\pi^{(+)}|^2+
(m_h^{(sf)})^2|\pi^{(+)}|^2.
\end{equation}
where $c_h^{(sf)}\sim{c}_v$ and $m_h^{(sf)}\sim\epsilon_v$. Notice that 
in general $c_h^{(sf)}\neq{c}_s^{(sf)}$ and $m_h^{(sf)}\neq{m}_s^{(sf)}$.
In particular we expect that $m_h^{(sf)}<m_s^{(sf)}$ in SFMFT because
antiferromagnetic correlation is weakened at the center of a vortex. 
As a result holes can hop more easily at vortex core and gain excess 
kinetic energy. Notice also that as in the case of spinons, the holons 
are also coupled to gauge fields $\vec{A}$ and $\vec{A}^s$. The precise 
way they are coupled to gauge fields depends on detail microscopic 
dynamics which cannot be obtained unambigously in our phenomenological 
analysis. We have ignored these couplings in writing down $L_{eff}^{h}$.

  It is interesting to compare the excitations described by the 
effective actions $L_{eff}^s$ and $L_{eff}^h$ with the corresponding
elementary excitations in d-wave superconducting state in SBMFT of 
$t-J$ model. First we consider the spin excitations. In both cases 
the spin excitations are described by pockets of Dirac fermions. The
number of species of Dirac fermions are the same in both cases -
there are four 'half-pockets' of Dirac-fermions centered around
$(k_x,k_y)\sim(\pm{k}_o,\pm{k}_o)$ in the d-wave superconducting state,
and there are two 'full-pockets' of Dirac fermions in $L_{eff}^s$ in 
order to respect parity. The position of the 'Dirac-pockets' in 
$\vec{k}$-space cannot be determined with certainty in $L_{eff}^s$.
Nevertheless, if we assume that the spinons are represented correctly
by $L_{eff}^s$ and the position of the fermion pockets are also 
centered around $(k_x,k_y)\sim(\pm{k}_o,\pm{k}_o)$, then the only
difference between spin excitations described by $L_{eff}^s$ and 
by SBMFT is that there is a gap $m_s^{(sf)}>0$ in the spinon spectrum 
of SFMFT. 

  The holon spectrum described by $L_{eff}^h$ is also similar to the 
elementary hole spectrum in SBMFT at momenta $\vec{k}\sim(0,0)$. The 
hole spectrum in SBMFT has dispersion\cite{sbmft}
\[
\epsilon(\vec{k})=-2t_{eff}(cos(k_x)+cos(k_y)), \]
where $t_{eff}\sim{t}$ and has energy minimum at $\vec{k}=(0,0)$. 
Around these regions the effective Lagrangians for holes have 
exactly the same form as \ (\ref{leffh}) in the non-relativistic 
limit with finite chemical potential $\mu=m_h^{(sf)}$ at $T=0$ and
$(c_h^{(sf)})^2/m_h^{(sf)}\sim{t}_{eff}$. The main differences between 
holons in SFMFT and holes in SBMFT are (1)the presence of nonzero 
chemical potential $\mu=m_h^{(sf)}$ in SBMFT which allows for
finite concentration of holes in ground state and (2)$(c_h^{(sf)})^2/
m_h^{(sf)}$ is expected to be of order $J$ in SFMFT. Notice that it 
has been argued that the effective hole band-width in SBMFT should be of 
order $\sim{J}$ after renormalization. In this case, the only 
qualitative difference between holon spectrum in SFMFT and hole 
spectrum in SBMFT is the appearance of nonzero mass gap and absence of 
Bose-condensation of holons in SFMFT. 

Lastly we discuss the effects of Bose-condensation of bosonic spins
($<Z>\neq0$) in the $T\rightarrow0$ limit. In this case the elementary
'bosons' which Bose-condense are single Schwinger bosons with (fictitious) 
charge one and vortices carrying $\pi$-flux quantum become unstable. 
The stable vortices carry $2\pi$-flux quantum. As a result the spinons 
and holons are confined in pairs. At finite temperatures the confining 
potential is effective up to length scale $\sim$ antiferromagnetic 
correlation length $\xi(T)$ and the effect of confinement is expected 
to be strong at low temperature when the system is at the 'renormalized
classical'\cite{chn} regime. As a result the identity of spinons and holons as 
independent excitations are lost at this regime of the phase dagram. A 
more detailed discussion of their experimental consequences is presented 
in section V and also in ref.\cite{ng2}.

\section{topological excitations in slave-boson mean-field theory}
  In SBMFT the spins are represented by fermion operators $c_{i\sigma}^+,
c_{i\sigma}$ and holes by boson operators $b^+_i,b_i$. The mean-field
Hamiltonian has the same form as the mean-field Hamiltonian in
SFMFT, $H_{MF}=H^s_{mf}+H^h_{mf}$, where\cite{sbmft}
\begin{eqnarray}
\label{sbmfs}
H^s_{mf} & = & -{J\over2}\sum_{<i,\nu>}(\Delta^*_{i\nu}(c_{i\uparrow}
c_{i+\nu\downarrow}-c_{i\downarrow}c_{i+\nu\uparrow})+H.C.)
\\    \nonumber
& & -\sum_{<i,j>,\sigma}\left({J\over2}\chi^*+t\delta)c^+_{i\sigma}
c_{j\sigma}+H.C.\right)+\sum_i\lambda\left(\sum_{\sigma}(c^+_{i\sigma}
c_{i\sigma}-(1-\delta)\right),
\end{eqnarray}
and
\begin{equation}
\label{sbmfh}
H^h_{mf}=-\sum_{<i,j>}\left(t\chi{b}^+_jb_i+H.C.\right)-\sum_i\mu
b^+_ib_i.
\end{equation}
The mean-field parameters are determined by the mean-field equations
$\Delta_{\nu}=<c_{i\uparrow}c_{i+\nu\downarrow}-c_{i\downarrow}
c_{i+\nu\uparrow}>$, $\chi=\sum_{\sigma}<c^+_{i\sigma}c_{j\sigma}>$,
and $\sum_{\sigma}<c^+_{i\sigma}c_{i\sigma}>=1-\delta$ and $<b^+_ib_i>
=\delta$. The spin-pairing parameter $\Delta_{\nu}$ has d-symmetry 
in d-wave superconducting state of SBMFT\cite{sbmft}.

   We shall follow the $SU(2)$ description of Wen {\em et.al.}\cite{wen4}
in describing the 'uniform' fluctuations of the order parameters 
$\Delta$ and $\chi$ in SBMFT. In this description, the low energy 
fluctuations of the system is described by an $O(3)$ order parameter 
$\vec{n}$, where $(n_1,n_2)=(Re\Delta_{\nu},Im\Delta_{\nu})$ describes 
the usual phase fluctuations of the d-wave spin-pairing order parameter, 
and $n_3$ corresponds to staggered flux fluctuations ($\chi$). At 
half-filling ($\delta=0$), the system has a $SU(2)$ pseudo-spin symmetry 
and the fluctuations described by $\vec{n}$ is $O(3)$ symmetric. The 
topological excitations are Skyrmions in this case. The $O(3)$ symmetry 
is broken down to $O(2)$ upon introduction of holes\cite{wen4} and the
fluctuations are described by an easy-plane anisotropic $O(3)$ model 
at finite doping with anisotropy energy of order $\delta^2J$\cite{su2}. 
The topological excitations become vortex-like half-Skyrmions with 
configuration $n_1\sim{n}cos\theta$, $n_2\sim{n}sin\theta$, $n_3\sim{0}$ 
(in a polar coordinate) away from center of vortex, and $n_3\sim{n}$ at 
center, i.e. instead of destroying the spin-correlations completely,
the vortex core is in the staggered-flux phase. The vortices carry
$\pi$-flux quantum as in usual superconductors. 

   Notice however that because holes exist as an independent quantum
liquid and can Bose-condense independently in SBMFT, two different 
kinds of vortices exist in SBMFT, corresponding to vortices associated 
with phase singularity in the d-wave spin-pairing order parameter 
$\Delta_{\nu}$, and vortices associated with the phase singularity of the 
hole condensate\cite{sa,nlv}. The two kinds of vortices carry different 
fluxes $\pi$ and $2\pi$,respectively\cite{sa,nlv} because of the 
different (fictitious) gauge charge associated with the spin pairs 
and holes. In the presence of hole condensate $<b>\neq0$, vortex 
excitations carrying fictitious flux $2\pi$ is the only stable
topological excitation\cite{vreal} and vortices carrying flux  
$\pi$ are confined, similar to the situation in SFMFT when $<Z>\neq0$. 
We shall consider a disordered state of holes where $<b>=0$ in the 
following. In this case, vortices with $\pi$ flux quantization are 
stable topological excitations. The effect of Bose-condensation will 
be addressed in next section when we discuss the $\delta-T$ phase 
diagram. Topological spin and charge excitations (spinons and holons)
can be constructed as in the case of SFMFT by considering bound 
states of spins and holes at the center of vortices. In particular, 
statistics transmutation occurs for particles binding to vortices 
with $\pi$ flux quantization, as in the case of SFMFT, i.e. we have
bosonic spinons and fermionic holons as topological excitations in
SBMFT when $<b>=0$.

   To study the dynamic of vortices we start with an effective $O(2)$ 
GL functional describing the phase fluctuations of the d-wave 
spin-ordering parameter\cite{su2},
\begin{equation}
\label{sbss}
S_{phase}={J\over2}\int{d}\tau\int{d}^2x\left[\kappa_x
(\nabla\theta-2\vec{A})^2+{2\delta\over{J}}
(\partial_{\tau}\theta-2A_{\tau})+\kappa_{\tau}
(\partial_{\tau}\theta-2A_{\tau})^2\right],
\end{equation}
where $\kappa_x\sim\delta$, $\kappa_{\tau}J\sim{J}$ is the 
compressibility of the spin-condensates, $\theta$ is the phase of 
the spin-pairing order parameter $\Delta$ and $A_{\mu}$ is the 
fictitious U(1) gauge field associated with phase fluctuations of the 
order parameter $\chi$. Notice the existence of linear 
$\partial_{\tau}$ term in $S_{phase}$ because of broken particle-hole symmetry\cite{su2}. The effective action describing dynamics of vortices 
can be constructed by following the same procedure as in last section. 
The resulting action $S_v$ has the same form as Eq.(8), except that 
the parameters $\xi_x$ and $\xi_{\tau}$ are replaced by $\kappa_x$ and 
$\kappa_{\tau}$, respectively and the presence of additional lnear
$\partial_{\tau}$ term in SBMFT. Correspondingly, the parameters 
$\epsilon_v$ and $c_v$ are modifed to $\epsilon_v\sim\kappa_xJ$ 
and $c_v\sim\sqrt{\kappa_x/\kappa_{\tau}}$. Notice that the linear
$\partial_{\tau}$ term does not introduce any strong effect on vortex 
dynamics because of screening of the vortex current $\vec{j}_v$ by 
$\vec{A}$ field, as in the case of SFMFT.

   The dynamics for the topological spin and charge excitations can
be obtained by making the same assumption as in last section that the
spacial degrees of freedom of the spins and holes are quenched once 
they are bound to the center of vortices. In particular, the dynamics 
of the spinons are described by a relativistic quantum field 
theory of charged bosons and the dynamics of the holons by a theory 
of Dirac fermions. We first consider the spinon excitations.
Following the same argument as in previous section, we construct two 
quantum fields $\psi^{(+)}_{\sigma}$ and $\psi^{(-)}_{\sigma}$ with
parity $\pm1$. Assuming that the low energy physics of spinons are 
described by the $\psi^{(+)}_{\sigma}$ field, we obtain for the spinons
\begin{equation}
\label{lefsbs}
L_{eff}^{s}=|\partial_{\tau}\psi^{(+)}_{\sigma}|^2+
|c_s^{(sb)}\nabla\psi^{(+)}_{\sigma}|^2+(m_s^{(sb)})^2|\psi^{(+)}_{\sigma}|^2.
\end{equation}
where $c_s^{(sb)}\sim{c}_v$ and $m_s^{(sb)}\sim\epsilon_v$. Notice that 
$L_{eff}^{s}$ has the same form as the effective Lagrangian for 
spinwaves in the disordered phase of the 
non-linear-$\sigma$-model or SFMFT\cite{sfmft,nlsm}.

  To construct dynamics for the holons we first construct a four-component
Dirac spinor field $\pi$, with
\begin{equation}
\label{flux}
\pi(x)=\left(\begin{array}{r}
\pi^{A}(x)  \\
\pi^{B}(x)
\end{array}\right)\; , \;\;\;
\pi^{A(B)}(x)=\left(\begin{array}{r}
\pi^{A(B)}_{1}(x) \\  
\pi^{A(B)}_{2}(x)  
\end{array}\right)
\end{equation}
where $\pi^{A(B)}$'s are two component fermion fields introduced to 
describe positive and negative energy solutions of the Dirac 
equation. In terms of $\pi$ the effective Lagrangian which 
transforms correctly under parity is\cite{ng2}
\begin{equation}
\label{lsbh}
L_{eff}^s={i\over2}\left[\bar{\pi}\gamma^0(\partial_{\tau}\pi)-
(\partial_{\tau}\bar{\pi})\gamma^0\pi+\bar{\pi}\vec{\gamma}.
(c_h^{(sb)}\nabla\pi)-(c_h^{(sb)}\nabla\bar{\pi}).\vec{\gamma}
\pi\right]-m_h^{(sb)}\bar{\pi}\pi,
\end{equation}
where $\gamma^{\mu}$'s are usual $4\times4$ Dirac matrices in $(2+1)d$ 
with $\mu=0,1,2$, $c_h^{(sb)}\sim{c}_v$ and $m_h^{(sb)}\sim\epsilon_v$. 
Notice that as in the case of spinons in SFMFT there must be two 
"Dirac pockets" of holes in order to respect parity. Notice also that 
$c_h^{(sb)}\neq{c}_s^{(sb)}$ and $m_h^{(sb)}\neq{m}_s^{(sb)}$ in general 
and we expect $m_h^{(sb)}>m_s^{(sb)}$ in the present case because vortex 
core is in the flux-phase where antiferromagnetic correlation is 
enhanced\cite{wen4,su2}. As a result it costs kinetic energy for 
hole to stay at the vortex core.

  Let us now compare the dynamics of spinons and holons in SBMFT with
elementary spin and charge excitations in SFMFT. First we consider spin
excitations. In both cases the dynamics of spin excitations are described
by relativistic complex scalar fields (spinwaves). In the case of SFMFT, 
the dispersion minimum of the spinwave is located around $\vec{q}=\pm(\pi
+q_{\delta},\pi+q'_{\delta})$ in the spiral states, where $q_{\delta}$ 
and $q'_{\delta}$ depends on the types of spiral state being formed. 
The location of dispersion minimum for spinons in SBMFT cannot be 
determined with certainty without further understanding of their 
microscopic dynamics. Nevertheless, if we assume that the 
dispersion minimum is located at the same $\vec{q}$-point as spinwaves
in SFMFT, then the only difference between spinons in SBMFT and spinwaves
in SFMFT is that there is a finite energy gap in the spinon spectrum 
at $T=0$. Similar correspondence also exists between holons in SBMFT and 
elementary hole excitations in SFMFT. In the spiral states of SFMFT the 
holes form fermi sea pockets around $\vec{k}\sim(\pm\pi/2,
\pi/2)$\cite{sfmft} in our choosen gauge, which can be interpreted
phenomenologically as two Dirac fermion pockets centered at the same 
momenta points and with finite chemical potential $\mu>m_h^{(sb)}$ such 
that particle-hole symmetry is broken and there exists nonzero density of
(fermionic) holes in the ground state. The interpretation is valid at 
low energy $\epsilon<<m_h^{(sb)}$. On the other hand, $\mu=0$ for holons 
in SBMFT, implying that the ground state holon density is zero and there 
exists a nonzero energy gap for holon excitations in SBMFT.

\section{duality picture and phase diagram of $t-J$ model}
    In the previous two sections we demonstrate that topological
spin and charge excitations exist as well defined excitations in 
the disordered phases $<Z>=0$, $<b>=0$ of slave-fermion and 
slave-boson mean-field theories of the $t-J$ model, respectively. 
Moreover, the topological spin and charge excitations in one 
mean-field ground state behave as "gapped" elementary excitations of the 
other mean-field state. In this section we shall explore this relation 
further by showing that a consistent phase diagram of the $t-J$ model 
can be obtained if we assume that the ground state in one mean-field 
theory is a quantum-disordered state of the other mean-field theory. 
The disordering effect is coming from "condensation" of topological 
excitations. We shall divide our discussions in two parts. In part I 
we shall study the effective actions of the topological excitations 
in the two mean-field theories and argue that quantum phase transitions 
to disordered states are natural outcomes of the effective actions 
when concentration of hole $\delta$ changes. Based on this result  
and the requirement that the phase diagram derived from the
quantum-disordering picture of SFMFT and SBMFT must be consistent 
with each other we shall construct in part II a plausible
phase diagram of $t-J$ model.

\subsection{effective actions for topological excitations and 
quantum-disordered phases}
    First we consider the case of SFMFT. To show that quantum-disordered
state of topological excitations is a natural outcome of our theory
we start with the phase action \ (\ref{glgauge}) which describes the
(uniform) phase fluctuations of the order parameter $\Delta$ in SFMFT. A
quantum-disordered phase is found if the phase $\theta$ is disordered
in the ground state. The action \ (\ref{spsf}) is in fact a 3D
$x-y$ model (coupled to gauge field) and is known to go through a
transition to a disordered phase when the coupling constant $g(\delta)
=\xi_x\xi_{\tau}$ is less than certain critical value $g_c$\cite{xy}. 
The mass gap of vortices go to zero as the system approaches the 
transition point from the ordered phase\cite{fl,nodon}, showing that 
the quantum phase transition is driven by condensation of vortex loops.
(Notice that we assume here that the gauge field $\vec{A}$ is not in the 
confining phase\cite{xy}). It is easy to see from Eq.\ (\ref{xis}) that
$g_{\delta}$ is a decreasing function of $\delta$ in SFMFT, as least 
for small $\delta$. It is therefore reasonable to assume that there 
exists a critical value of hole concentration $\delta_c$ at which the 
ground state described by SFMFT goes through a quantum phase transition 
to a disordered phase when $\delta>\delta_c$. 

   The main difference between the quantum-disordered state in SFMFT 
and usual 3D $x-y$ model is that we assume a bound state of spin or 
hole exists in the vortex core in SFMFT. Thus there exists two kinds 
of vortices (spinons and holons) in SFMFT which possess different 
statistics and dynamics. In particular, since we assume
spin-charge separation and the mass gap for holons $m_h^{(sf)}$ should 
be smaller than the mass gap for spinons $m_s^{(sf)}$ in general, the
condensation of (bosonic) holons and (fermionic) spinons should occur 
at two different critical hole concentrations $\delta_{ch}^{(sf)}$ 
and $\delta_{cs}^{(sf)}$, respectively with $\delta_{ch}^{(sf)}
<\delta_{cs}^{(sf)}$. Therefore two different quantum-disordered states 
corresponding to condensation of holons and holons $+$ spinons are 
expected to exist in our theory. Unfortunately we cannot deduce the 
properties of the quantum-disordered states unambigously based 
on our simple phenomenological treatment of spinon and holon dynamics. 
In particular, the nature of the quantum-disordered state coming from
condensation of {\em fermionic} vortex loops has never been 
investigated theoretically. The nature of the quantum disordered states
will be deduced phenomenologically in the next subsection. Another 
special feature of SFMFT which is absent in usual 3D $x-y$ model is 
the presence of long-ranged spiral magnetic order $<Z>\neq0$ at zero
temperature, at least when hole concentration is small. We shall assume 
that the long-ranged order exists at small value of $\delta$ and 
vanishes ($<Z>\rightarrow0$) at exactly $\delta_{cs}^{(sf)}$, for 
reasons we shall see in the following. To summarize we show in fig.2a
schematically the expected behavior of $m_h^{(sf)}$, $m_s^{(sf)}$ and 
$<Z>$ as functions of $\delta$ in SFMFT.   

   Next we consider the case of SBMFT. We start with the phase 
action \ (\ref{sbss}) describing the phase dynamics of the d-wave 
spin-pairing order parameter $\Delta$ in SBMFT. The phase 
action \ (\ref{sbss}) is also a 3D $x-y$ model as in the case of 
SFMFT and the system is expected to go through a transition to a 
disordered phase when the coupling constant $g(\delta)=\kappa_x
\kappa_{\tau}$ is less than certain critical value $g_c$. It is easy 
to see that $g(\delta)$ is an increasing function of $\delta$ in 
SBMFT, at least for small $\delta$ and goes to zero as 
$\delta\rightarrow0$. Therefore, it seems reasonable to assume that 
there exists a critical value of hole concentration $\delta_c$ below 
which the system is in a quantum-disordered phase of SBMFT. 
 
  As in SFMFT, the main difference between the quantum-disordered 
state of SBMFT and usual 3D $x-y$ model (in the context of a d-wave
superconductor\cite{nodon}) is that we assume bound state of 
spin or hole exists in the vortex core and there are two kinds of 
vortices (spinons and holons) in SBMFT. Unlike SFMFT, we expect that 
the mass gap for holons $m_h^{(sb)}$ is {\em larger} than mass gap for 
spinons $m_s^{(sb)}$ in SBMFT and therefore, the condensation of 
(fermionic) holons and (bosonic) spinons should occur at different 
critical hole concentrations $\delta_{ch}^{(sb)}$ and $\delta_{cs}^{(sb)}$, 
respectively with $\delta_{ch}^{(sb)}<\delta_{cs}^{(sb)}$. Notice that 
we are not able to deduce the properties of the quantum-disordered states 
of SBMFT with certainty as in the case of SFMFT.

   Note that because of spin-charge seperation, the holes in SBMFT
may exist in a quantum disordered (insulator) phase with $<b>=0$ by 
themselves, independent of the disordering effect on spin-pairing order
parameter $\Delta$\cite{kim}. In particular, it is expected that a zero
temperature disordered phase of bosonic holes can be formed at small
enough $\delta<\delta_b^{(sb)}$ because of disordering effect associated 
with ($2\pi$ flux) vortex-loop condensation in the boson field\cite{kim}. 
The different disordering effects in SBMFT are summarized in fig.2b,
where we show schematically the expected behaviours of mass gap 
$m_h^{(sb)}$, $m_s^{(sb)}$ and hole-condensation amplitude $<b>$ as 
functions of $\delta$. Notice that we have assumed that $\delta_b^{(sb)}>
\delta_{cs}^{(sb)}$ in drawing the figure. We shall explain why 
we make this assumption in the following subsection.

\subsection{a plausible phase diagram of $t-J$ model}
   In this subsection we shall construct a phase-diagram of $t-J$
model, based on what we have obtained in previous sections. Our main
assumption is that the quantum-disordering effect associated with
condensation of topological excitations in SFMFT (SBMFT) of $t-J$ 
model gives rise to the ground state described by SBMFT (SFMFT), 
with the roles of topological excitations and elementary excitations 
being interchanged at the quantum-disordering transition. This
assumption is consistent with the "dual" relations we obtained in
sections II and III between the elementary excitations in one 
mean-field phase and topological excitations in the other. Notice 
that this statement has to be understood with care because the 
quantum phase transitions associated with condensation of spinons and 
holons occur in general at different dopping concentration 
$\delta$'s(see figure 2). As a result there exists at least three 
different regions in the zero-temperature phase diagram  of $t-J$ model
in the duality picture.

   For our assumption to be valid, the quantum-disordering transition
of holons and spinons from SFMFT to SBMFT and from SBMFT to SFMFT must 
occur at consistent value of doping, i.e. we must have
$\delta_{cs}^{(sf)}=\delta_{cs}^{(sb)}=\delta_{cs}$ and
$\delta_{ch}^{(sf)}=\delta_{ch}^{(sb)}=\delta_{ch}$. The zero-temperature 
phase diagram has therefore at least three different regions:
(i)$\delta<\delta_{ch}$, in this region the low energy physics is
described by SFMFT, fermionic spinons and bosonic holons are high-energy
topological excitations, (ii)$\delta_{ch}<\delta<\delta_{cs}$, in
this region we expect that both low energy spin and charge excitations
are bosonic. The precise nature of this state will be discussed below, 
and (iii)$\delta_{cs}<\delta$, in this region the low energy physics 
is described by SBMFT, bosonic spinons and fermionic holons are 
high-energy topological excitations.

   Next we consider the transition regions $\delta\sim\delta_{ch}$ 
and $\delta\sim\delta_{cs}$ more carefully. First we consider
$\delta\sim\delta_{ch}$. The change in the hole spectrum around this 
region can be described phenomenologically by a band picture with 
two branches of hole spectrum. The first one is the Dirac fermion 
spectrum \ (\ref{lsbh}) with Dirac pockets centered at $\vec{k}\sim(\pm\pi/2,\pi/2)$ and the second one is the boson 
spectrum \ (\ref{leffh}) with dispersion minimum around 
$\vec{k}\sim(0,0)$. In the regime $\delta<\delta_{ch}$, we have
$m_h^{(sf)}>\mu>m_h^{(sb)}$ such that at zero temperature, the Dirac
fermion pockets are occupied and the bosonic hole band is empty.
The ground state of the system is described by SFMFT. As 
$\delta\rightarrow\delta_{ch}$, $m_h^{(sf)}\rightarrow\mu$ and the
bosonic hole excitations become gapless. As $\delta$ increases,
$m_h^{(sf)}=\mu$ and decreases further, until all the fermionic holes 
become bosonic and the Dirac pocket becomes empty. Notice that this 
band picture describes only the transformation of fermionic holes to 
bosonic ones at $\delta\sim\delta_{ch}$ but the nature of the ground 
state of the bosonic hole system is not addressed. We shall argue below 
that the bosonic holes must exist in an insulating phase and 
$\delta_{ch}$ is a critical point where the system undergoes a 
metal-insulator transition in the absence of potential disorder 
(Anderson localization). Notice that this is consistent with the 
existence of a nonzero critical hole density $\delta_b^{(sb)}$ below 
which the bosonic holes are in a quantum-disordered phase\cite{kim}.

   Next we consider the region $\delta\sim\delta_{cs}$. The transition
can be described by a picture with two branches of spin excitations. 
The first one is the Dirac fermion spectrum \ (\ref{lspinon}) and 
the second one is the spinwave like spectrum \ (\ref{lefsbs}). In the 
regime $\delta< \delta_{cs}$, $m_s^{(sb)}$ is zero whereas $m_s^{(sf)}$ 
is finite, the presence of gapless spinwave excitation indicates that 
the system is magnetically ordered and the Dirac fermions are high 
energy spin excitations. Therefore we expect $<Z>\neq0$ at 
$\delta<\delta_{cs}$, as we have assumed in previous subsection 
(see fig.2a). The opposite is true for $\delta>\delta_{cs}$, where 
$m_s^{(sb)}$ becomes nonzero and $m_s^{(sf)}$ is zero. In this region
the elementary spin excitations are described by gapless Dirac fermion
pockets. The spinwave excitations are 'gapped' and the system has no 
long-ranged magnetic order. Notice that the system may be in a 
spin-gap' state or d-wave superconducting state, depending on whether
$<b>$ is equal to zero\cite{sbmft,ul}.

   To complete our discussion we shall now consider the regime
$\delta_{ch}<\delta<\delta_{cs}$ and ask what is the ground 
state of the system at this regime. Following our previous
discussions this is a regime where the spin dynamics are described
by SFMFT and hole dynamics by SBMFT with both $<Z>\neq0$ and $<b>\neq0$
in a naive mean-field picture. However, as pointed out in last section, 
this state cannot exists because $<b>\neq0$ implies that the (bosonic) 
topological spin excitations are confined and as a result we cannot 
have both $<Z>\neq0$ and $<b>=0$ in the ground state. Therefore,
we must have $<b>=0$ in order to have a consistent phase diagram, 
i.e. we expect that the bosonic holes are in a disordered (insulator) 
state at this regime of the phase diagram. As a result we expect
$\delta_b^{sb}\geq\delta_{cs}$, as is assumed in the previous
subsection (fig.(2b)). 

   The general $\delta-T$ phase diagram based on our duality picture
is shown in fig.3. We note that there are four different low
temperature regions in the phase diagram. Region I is the 'renormalized 
classical' regime of SFMFT. In this regime the (topological) fermionic 
spin and bosonic charge excitations are gapped and confined. In the 
absence of potential disorder, long-ranged (spiral) magnetic ordering is 
expected to exist in this regime when interlayer coupling is included. 
Region II is a new regime coming from the duality picture. In this 
regime the holes are in the Mott insulator state but long-ranged 
magnetic ordering still exists. Notice that the nature of the magnetic 
ordering is unclear in this regime. In SFMFT, spiral ordering comes 
as a result of formation of fermionic hole pockets\cite{sfmft} at 
momentum points $\vec{k}\sim(\pi/2,\pi/2)$. However, the hole pockets 
are gradually destroyed at this region of the phase diagram and we 
expect that long-range spiral ordering should also be destroyed. It 
is known experimentally that the effect of potential disorder is very 
important in the antiferromagnetic and spin-glass regime of high-$T_c$ 
cuprates\cite{rmp}. In particular, it seems that the holes in the
antiferromagnetic phase are always localized by disorder and long-range
spiral ordering never appears\cite{rmp}. It is expected that the potential
disorder will also affect the magnetic behaviour at region II strongly 
and the precise nature of the ground state is unclear. In particular,
a spin-glass like phase may appear if the localized holes are mobile
enough at short distance scale so that random long-ranged 
spin-distortions are set up in the system\cite{gooding}. Region III is 
the spin-gap phase described by SBMFT where the low energy spin 
excitations are described by pockets of Dirac fermions and the 
antiferromagnetic (spiral) spinwaves are gapped. The holes are still in 
the Mott insulator state. As a result the (topological) bosonic spin 
and fermionic charge excitations in this regime are gapped but 
{\em deconfined}. Region IV is the $<b>\neq0$ state. The ground state 
is a d-wave superconductor described by SBMFT. The topological excitations 
are {\em confined} since $<b>\neq0$. Corresponding to the 
zero-temperature quantum phase transitions between these four phases are 
three different quantum critical regimes (a-c) indicated on the 
phase-diagram. Note that region III and the corresponding quantum 
critical regime may vanish if $\delta_{cs}=\delta_b^{sb}$ which is allowed 
in the duality picture. When comparing with the experiments on 
monolayer copper oxides\cite{rmp} we find that the gross features of the 
phase diagrams can be produced in the duality picture. Moreover, 
new branches of excitations not present in mean-field theories are 
predicted in the duality picture. A few experimental consequences from 
the duality picture will be discussed in the next section where our 
theory will be critically examined. 

\section{summary and discussions}
   There are several new results obtained in this paper. In sections II 
and III, we showed that topological spin and hole excitations exist in
general in the disorder states of SFMFT and SBMFT, and the dynamics 
of these excitations can be derived under some very general assumptions. 
We found that the topological spin and charge excitations in one 
mean-field theory have same dynamics as elementary excitations in the 
other mean-field theory, except the appearance of energy gaps. Based
on this result, we propose in section IV a duality relation between the
ground states described by the two mean-field theories, namely that
the ground state described by one mean-field theory can be viewed as a
quantum-disordered state of the other mean-field theory.
A phase diagram of $t-J$ model is constructed phenomenologically based
on this assumption. In particular we find that the self-consistent
requirement between the quantum-disordering effects from SFMFT and
SBMFT implys that an intermediate insulator phase between the 
antiferromagnetic (spiral) and d-wave superconducting phases has to exist.
In the following subsections we shall discuss some theoretical questions 
and experimental consequences of our theory. We first consider the 
theoretical questions.

\subsection{theory}
   To have some qualitative understanding on the meaning of the duality 
relation we shall now give a wavefunction interpretation of the relation. 
First we consider the spin part of the ground state wavefunction described 
by SFMFT. As pointed out in section II, the wavefunction corresponding to 
SFMFT is a RVB state, where the wavefunction is a superposition of 
spin-singlet pairs. The relative phases between the singlet pairs are fixed
in SFMFT. The spin part of the ground state wavefunction described by 
SBMFT has a similar interpretation. The wavefunction is also made up of
superposition of spin-singlet pairs, and the relative phases between the 
singlet pairs are fixed by SBMFT. The difference between SFMFT and SBMFT
is that because of the different particle statistics used in representing
the spins, the phase relations between singlet pairs are very different in
the two mean-field theories\cite{rc}. The differences in phase relations
between spin-singlet pairs result in the very different physical 
properties represented by the two RVB wavefunctions. Our duality
relation suggests a connection between the two very different RVB 
wavefunctions, namely, the phase relation between spin singlets in one
wavefunction can be obtained from quantum-disordering the phases of the 
other wavefunction through condensation of spinons and holons. Notice that 
a natural consequence of this picture is that there should be no clear 
distinction between a high temperature short-ranged antiferromagnetic 
state and preformed Cooper pair state in $t-J$ model when the phases 
between the RVB singlets pairs are randomized. Differences between 
the two phases show up only as temperature is lowered and phase 
coherences between RVB pairs are built up gradually.

  Notice that the effects of holes is not mentioned in the above 
picture. In our derivation of the duality relation, we have assumed a
picture of spin-charge separation where spins and holes can be 
considered as independent quantum liquids. This is also a major 
assumption we made in deriving our phase diagram. Although it is believed
that the picture of spin-charge separation should be correct at an
intermediate temperature region above $T_c$, it is also believed that
a proper treatment of the effect of (fictitous) gauge fields may
lead to confinement of spin and charge at low enough temperature in 
$t-J$ model\cite{wl,nl1} and may modify strongly our theoretical 
predictions. In particular, region III of our phase diagram may be washed 
away completely if spin and charge are confined by fictitious gauge
fields at low energy. The effects of gauge fields are ignored 
in our present theory which treats the dynamics of topological
spin and charge excitations only in a semi-phenomenological level. 
To include the effects of gauge fields properly a more microscopic 
understanding of topological excitations is needed and will be a 
direction of our future work\cite{ngsuper}.

\subsection{experimental consequences}
   Next we consider a few experimental consequences of the duality 
relations. First we consider regions I and IV where the ground
states are described by SFMFT and SBMFT, respectively. The new
features we predict in these two regimes are the existence of
topological spin and charge excitations which are absent in mean-field
theories. Notice that as discussed in sections II and III, the 
topological excitations are {\em confined} in these regimes and are
difficult to be observed in correlation functions with momentum
transfer $\vec{q}\sim(0,0)$. Nevertheless, the behavior of spin
correlation functions at momentum $\vec{q}\sim(\pi,\pi)$ is 
determined by the short-distance behaviours of spin {\em pairs} and 
should not be affected strongly by confinement. As a result, we 
expect that topological spin excitations may be observable in the 
dynamic structure factor $S(\vec{q},\omega)$ at momenta
$\vec{q}\sim(\pi,\pi)$\cite{ng2}. A more detailed analysis of the 
effects of topological spin excitations in the $\delta=0$ limit 
(Heisenberg model) was given in Ref.\cite{ng2}
(see also Ref.\cite{ngsuper}). Notice that gapped 
incommensurate spinwave excitations with momentum $\vec{q}\sim
(\pi\pm2\pi\delta,\pi)$ and $(\pi,\pi\pm2\pi\delta)$ have been 
observed experimentally in the underdoped regime of high-$T_c$
cuprates\cite{incs} which we believe, can be identified with the 
bosonic topological spin excitations in the d-wave superconductor 
phase. Notice that the topological excitations can be observed 
more easily in the underdoped regime at $T>T_c$ or at the spin-gap 
phase (region III) where confinement effect is absent. The spin-gap 
phase has low-energy spin dynamics of d-wave superconductors but is 
insulating otherwise. It would be interesting to see whether this 
phase can be observed in cuprates.  

  Lastly we discuss region II. This is a regime where our understanding
is poorest. Based on our analysis we expect that the system is in an
insulator phase where charge excitations are gapped, and the spins are
in some kind of long-ranged order state. However, the nature of the
spin-ordering is unclear, especially in the presence of impurities. We
expect that this phase may corresponds to the spin-glass 
phase\cite{rmp} observed experimentally. Theoretical understanding
of spin dynamics at this regime is still very limited\cite{gooding}. 
Notice that it is found experimentally that the in-plane resistance of 
the cuprates at the spin-glass regime goes as $\rho_{ab}(T)\sim{k}_BT$ 
at high enough temperature, before the systems become 
insulating\cite{tak}. Within our duality picture, a plausible 
explanation is that the low energy hole excitations are already 
slave-boson-like in the spin-glass phase. The transport behaviours of 
the systems at high enough temperature should be described by SBMFT, 
which predicts $\rho_{ab}(T)\sim{k}_BT$ at high enough temperature 
because of gauge field fluctuations.
  
  The author thanks N. Nagaosa, P.A. Lee, Y.B. Kim and Z.B. Su for
many helpful discussions. This work is support by HKRGC through 
Grant no. HKUST6124-98P.

\begin{figure}
\caption{vortex excitations in SFMFT of $t-J$ model; (1a)S=0, chargeless 
vortex excitation. The center of vortex is located at the center of a
square plaquette; (1b)spinon; (1c)holon. Notice that the center of vortex
is located at a lattice site for both spinon and holon. Similar 
constructions can also be applied to the case of SBMFT.}
\end{figure}

\begin{figure}
\caption{(2a)Schematic behavior of $m_h^{(sf)}$, $m_s^{(sf)}$ and $<Z>$ as
functions of hole concentration $\delta$ in SFMFT; line (a): $m_s^{(sf)}$,
line (b): $m_h^{(sf)}$, blacked region: $<Z>\neq0$; (2b)Schematic behavior
of $m_h^{(sb)}$, $m_s^{(sb)}$ and $<b>$ as functions of hole concentration
$\delta$ in SBMFT; line (a): $m_s^{(sb)}$, line (b): $m_h^{(sb)}$, blacked
region: $<b>\neq0$.}
\end{figure}

\begin{figure}
\caption{phase diagram of $t-J$ model. Region I: renormalized classical
regime of quantum antiferromagnet described by SFMFT. Region II: spin-glass
regime with spin dynamics described by SFMFT and hole dynamics by SBMFT.
Region III: spin-gap regime described by SBMFT with $<b>=0$. Region IV:
d-wave superconductor phase described by SBMFT. Regions (a)-(c) are 
the corresponding quantum critical regimes of the zero temperature
phase transitions}
\end{figure}

\begin{references}
\bibitem{vreal} Notice that this is correct only for vortices binding
   to fictitous gauge flux. See also refs.[17,18].
\end{references}
\end{document}